\documentstyle[prd,epsfig,aps,amsmath,floats,tighten,doublespace]{revtex}

\newlength{\capwidth}
\newcommand{\ttbar}{$t\bar{t}$}

\newcommand{\MET}{\rlap{\kern0.25em/}E_T}

\begin{document}


\begin{center}
   {\Large\bf Search for $ZZ$ and $ZW$ production in $p\bar{p}$ collisions \\
	at $\sqrt{s}$ = 1.96 TeV}\\
   \vspace*{1.0cm}
   \font\eightit=cmti8
\def\r#1{\ignorespaces $^{#1}$}
\hfilneg
\begin{sloppypar}
\noindent 
D.~Acosta,\r {16} J.~Adelman,\r {12} T.~Affolder,\r 9 T.~Akimoto,\r {54}
M.G.~Albrow,\r {15} D.~Ambrose,\r {43} S.~Amerio,\r {42}  
D.~Amidei,\r {33} A.~Anastassov,\r {50} K.~Anikeev,\r {15}
% ...
\end{sloppypar}
\vskip .026in
\begin{center}
(CDF Collaboration)
\end{center}

\vskip .026in
\begin{center}
\r 1  {\eightit Institute of Physics, Academia Sinica, Taipei, Taiwan 11529, 
Republic of China} \\
\r 2  {\eightit Argonne National Laboratory, Argonne, Illinois 60439} \\
\r 3  {\eightit Institut de Fisica d'Altes Energies, Universitat Autonoma
de Barcelona, E-08193, Bellaterra (Barcelona), Spain} \\
\r 4  {\eightit Istituto Nazionale di Fisica Nucleare, University of Bologna,
I-40127 Bologna, Italy} \\
\r 5  {\eightit Brandeis University, Waltham, Massachusetts 02254} \\
\r 6  {\eightit University of California at Davis, Davis, California  95616} \\
\r 7  {\eightit University of California at Los Angeles, Los 
Angeles, California  90024} \\
\r 8  {\eightit University of California at San Diego, La Jolla, California  92093} \\ 
\r 9  {\eightit University of California at Santa Barbara, Santa Barbara, California 
93106} \\ 
\r {10} {\eightit Instituto de Fisica de Cantabria, CSIC-University of Cantabria, 
39005 Santander, Spain} \\
\r {11} {\eightit Carnegie Mellon University, Pittsburgh, PA  15213} \\
\r {12} {\eightit Enrico Fermi Institute, University of Chicago, Chicago, 
Illinois 60637} \\
\r {13}  {\eightit Joint Institute for Nuclear Research, RU-141980 Dubna, Russia}
\\
\r {14} {\eightit Duke University, Durham, North Carolina  27708} \\
\r {15} {\eightit Fermi National Accelerator Laboratory, Batavia, Illinois 
60510} \\
\r {16} {\eightit University of Florida, Gainesville, Florida  32611} \\
\r {17} {\eightit Laboratori Nazionali di Frascati, Istituto Nazionale di Fisica
               Nucleare, I-00044 Frascati, Italy} \\
\r {18} {\eightit University of Geneva, CH-1211 Geneva 4, Switzerland} \\
\r {19} {\eightit Glasgow University, Glasgow G12 8QQ, United Kingdom}\\
\r {20} {\eightit Harvard University, Cambridge, Massachusetts 02138} \\
\r {21} {\eightit The Helsinki Group: Helsinki Institute of Physics; and Division of
High Energy Physics, Department of Physical Sciences, University of Helsinki, FIN-00044, Helsinki, Finland}\\
\r {22} {\eightit Hiroshima University, Higashi-Hiroshima 724, Japan} \\
\r {23} {\eightit University of Illinois, Urbana, Illinois 61801} \\
\r {24} {\eightit The Johns Hopkins University, Baltimore, Maryland 21218} \\
\r {25} {\eightit Institut f\"{u}r Experimentelle Kernphysik, 
Universit\"{a}t Karlsruhe, 76128 Karlsruhe, Germany} \\
\r {26} {\eightit High Energy Accelerator Research Organization (KEK), Tsukuba, 
Ibaraki 305, Japan} \\
\r {27} {\eightit Center for High Energy Physics: Kyungpook National
University, Taegu 702-701; Seoul National University, Seoul 151-742; and
SungKyunKwan University, Suwon 440-746; Korea} \\
\r {28} {\eightit Ernest Orlando Lawrence Berkeley National Laboratory, 
Berkeley, California 94720} \\
\r {29} {\eightit University of Liverpool, Liverpool L69 7ZE, United Kingdom} \\
\r {30} {\eightit University College London, London WC1E 6BT, United Kingdom} \\
\r {31} {\eightit Massachusetts Institute of Technology, Cambridge,
Massachusetts  02139} \\   
\r {32} {\eightit Institute of Particle Physics: McGill University,
Montr\'{e}al, Canada H3A~2T8; and University of Toronto, Toronto, Canada
M5S~1A7} \\
\r {33} {\eightit University of Michigan, Ann Arbor, Michigan 48109} \\
\r {34} {\eightit Michigan State University, East Lansing, Michigan  48824} \\
\r {35} {\eightit Institution for Theoretical and Experimental Physics, ITEP,
Moscow 117259, Russia} \\
\r {36} {\eightit University of New Mexico, Albuquerque, New Mexico 87131} \\
\r {37} {\eightit Northwestern University, Evanston, Illinois  60208} \\
\r {38} {\eightit The Ohio State University, Columbus, Ohio  43210} \\  
\r {39} {\eightit Okayama University, Okayama 700-8530, Japan}\\  
\r {40} {\eightit Osaka City University, Osaka 588, Japan} \\
\r {41} {\eightit University of Oxford, Oxford OX1 3RH, United Kingdom} \\
\r {42} {\eightit University of Padova, Istituto Nazionale di Fisica 
          Nucleare, Sezione di Padova-Trento, I-35131 Padova, Italy} \\
\r {43} {\eightit University of Pennsylvania, Philadelphia, 
        Pennsylvania 19104} \\   
\r {44} {\eightit Istituto Nazionale di Fisica Nucleare, University and Scuola
               Normale Superiore of Pisa, I-56100 Pisa, Italy} \\
\r {45} {\eightit University of Pittsburgh, Pittsburgh, Pennsylvania 15260} \\
\r {46} {\eightit Purdue University, West Lafayette, Indiana 47907} \\
\r {47} {\eightit University of Rochester, Rochester, New York 14627} \\
\r {48} {\eightit The Rockefeller University, New York, New York 10021} \\
\r {49} {\eightit Istituto Nazionale di Fisica Nucleare, Sezione di Roma 1,
University di Roma ``La Sapienza," I-00185 Roma, Italy}\\
\r {50} {\eightit Rutgers University, Piscataway, New Jersey 08855} \\
\r {51} {\eightit Texas A\&M University, College Station, Texas 77843} \\
\r {52} {\eightit Texas Tech University, Lubbock, Texas 79409} \\
\r {53} {\eightit Istituto Nazionale di Fisica Nucleare, University of Trieste/\
Udine, Italy} \\
\r {54} {\eightit University of Tsukuba, Tsukuba, Ibaraki 305, Japan} \\
\r {55} {\eightit Tufts University, Medford, Massachusetts 02155} \\
\r {56} {\eightit Waseda University, Tokyo 169, Japan} \\
\r {57} {\eightit Wayne State University, Detroit, Michigan  48201} \\
\r {58} {\eightit University of Wisconsin, Madison, Wisconsin 53706} \\
\r {59} {\eightit Yale University, New Haven, Connecticut 06520} \\
\end{center}

\vspace*{0.5cm}

(Dated: \today)

\vspace*{0.5cm}

\parbox{\capwidth}
{\rightskip=0.5cm\leftskip=0.5cm\noindent\baselineskip=12pt
We present a search for $ZZ$ and $ZW$ vector boson pair production in $p\bar{p}$ 
collisions at $\sqrt s$ = 1.96 TeV using the leptonic decay channels 
$ZZ\rightarrow ll\nu\nu,~ZZ\rightarrow lll'l'$ 
and $ZW\rightarrow lll'\nu$.
In a data sample corresponding to an integrated 
luminosity of 194 pb$^{\mathrm -1}$ collected with the Collider Detector 
at Fermilab, 3 candidate events are found with an expected background of
$1.0 \pm 0.2$ events. We set a 95\% confidence level upper limit of 15.2 pb on the cross 
section for $ZZ$ plus $ZW$ production, compared to the standard model prediction of 
5.0 $\pm$ 0.4 pb.

\vspace*{0.5cm}

}
\end{center}

\vspace*{1.0cm}

\twocolumn

The measurements of $ZZ$ and $ZW$ production provide a direct test of the 
standard model (SM) prediction of triple--gauge--boson couplings~\cite{TGC}.
The presence of unexpected neutral triple--gauge--boson couplings ($ZZZ$ and $ZZ\gamma$)
can result in an enhanced rate of $ZZ$ production, and an anomalous $WWZ$ coupling
can increase the $ZW$ production rate above the SM prediction. 
The $WWZ$ and $ZZ\gamma$ couplings have been studied by the CDF and D\O\ experiments through 
the study of $WW$, $ZW$, $W\gamma$, and $Z\gamma$ production [2-8].   
The D\O\ experiment has measured an upper limit on the cross section for $ZW$ production~\cite{d0TGC2}. 
No limit has been set on the cross section for $ZZ$ production from hadron collisions, 
but production properties have been studied at LEP\,II in $e^+ e^-$ collisions at 
$\sqrt{s} = 183 - 209$ GeV~\cite{lepZZ}.  A comprehensive review of the limits on 
anomalous $WWZ$, $ZZZ$, and $ZZ\gamma$ couplings at LEP\,II can be found in Ref.~\cite{lepTGC}.
The production of $WZ$ and $ZZ$ boson pairs is also of interest because the decays cause 
significant backgrounds in searches for the SM Higgs boson.

In this report we present a search for $ZZ$ and $ZW$ production using the three decay modes
$ZZ\rightarrow ll\nu\nu$, $ZZ\rightarrow lll'l'$, and $ZW\rightarrow lll'\nu$,
where $l$ and $l'$ are electrons or muons, 
predominantly from direct $W$ or $Z$ decays, but also with a small contribution 
from the leptonic decay of tau leptons. This study is based on 
($194 \pm 12)$~pb$^{-1}$~\cite{lum} of data collected by the upgraded Collider 
Detector at Fermilab (CDF\,II) from March 2002 to September 2003 using $p\bar{p}$ collisions 
at $\sqrt s$ = 1.96~TeV.
CDF\,II is a general-purpose detector at the Tevatron accelerator at Fermilab. 
The main components used in this analysis are a silicon 
vertex detector, a central tracking drift chamber, 
central ($|\eta| < 1.1$~\cite{xyz}) and forward ($1.1<|\eta|<3.6$)
electromagnetic and hadronic calorimeters, and muon chambers. 
The silicon detector and central tracking chamber are located inside a 1.4~T superconducting
solenoidal magnet. A more detailed description of the detector can be
found in the CDF technical design report~\cite{CDF3} and in a recent publication describing
a measurement of $WW$ pair production~\cite{cdfTGC3}.

The data samples are collected by a trigger system that selects events having electron candidates 
in the central calorimeter with $E_T >$ 18 GeV, or muon candidates with $p_T >$ 18 GeV/$c$.
Events for this analysis are then selected by requiring at least two leptons with 
$E_T >$ 20 GeV and $|\eta| < 2.5$ for electrons, or $p_T >$ 20 GeV/$c$ and $|\eta| < 1$ for muons.
An electron is identified as energy deposited in the central electromagnetic calorimeter which is 
matched to a well-measured track reconstructed in the central tracking chamber, 
or, for an electron with $|\eta| > 1.2$, as energy deposited in the forward electromagnetic 
calorimeter with an associated track utilizing a calorimeter-seeded silicon
tracking algorithm~\cite{phxTrk}. 
In addition, electrons must have appropriate shower profiles in the electromagnetic calorimeters.
A muon is identified as a track in the central tracking chamber, with
energy deposition in the calorimeter consistent with a minimum ionizing particle,
and with a track segment in the muon chambers. 
If a minimum ionizing track points towards a gap in the muon chamber coverage, 
it is still considered a muon candidate in events that have an additional 
electron in the central calorimeter or a muon with a muon chamber track segment.
All charged leptons are required to be isolated from additional nearby calorimeter activity.
The transverse energy deposited around an electron or muon in a cone of radius 
$\Delta R = \sqrt {\Delta\phi^2 + \Delta\eta^2} = 0.4$, excluding the calorimeter energy 
matched to the lepton candidate, is required to be less then 
10\% of the electron $E_T$ or muon $p_T$.

The signature of neutrinos in the decays of $ZZ\rightarrow ll\nu\nu$ 
and $ZW\rightarrow lll'\nu$ is missing transverse energy ($\MET$),
measured from the imbalance of $E_T$ in the calorimeter
and the escaping muon $p_T$ (when muon candidates are present).

The next-to-leading-order (NLO) $ZZ$ and $ZW$ cross sections in $p\bar{p}$ collisions 
at $\sqrt s$ = 1.96 TeV are calculated by the MCFM~\cite{mcfm} 
program using the CTEQ6 parton distribution functions~\cite{pdf}. 
They are $\sigma(p\bar p \rightarrow ZZ)$ = ($1.39 \pm 0.10$) pb
and $\sigma(p\bar p \rightarrow ZW)$ = ($3.65 \pm 0.26$) pb. 

We study events in three categories designed to encompass the main leptonic
branching ratios of the $ZZ$ and $ZW$ decays. 
The first includes events with four charged leptons,
which is sensitive to $ZZ\rightarrow lll'l'$ ($l$ and $l'=e,\mu,\tau$) with
a branching ratio of 1.0\%.
Since we only select events with electrons and/or muons, we are only 
sensitive to final--state taus through their subsequent decay to leptons.
The second category, which includes events with three charged leptons plus $\MET$, 
consists predominantly of $ZW\rightarrow lll'\nu$
(branching ratio of 3.3\%). Events from $ZZ\rightarrow lll'l'$, where one lepton 
is not identified, can also fall into this category. 
The third category includes events with two charged leptons 
plus $\MET$, which is sensitive to $ZZ\rightarrow ll\nu\nu$ (branching ratio of 4.0\%) 
and $ZW\rightarrow lll'\nu$, where one lepton is not identified.

Our strategy is to first select events containing a $Z$ boson and then 
require additional leptons and/or large $\MET$ in the event. 
The $Z$ boson is identified by one pair of same-flavor oppositely-charged 
leptons ($e^+ e^-$ or $\mu^+ \mu^-$) with an invariant mass between 76 GeV/$c^2$ and 106 GeV/$c^2$. 
The four-lepton category selects a second lepton pair using the same criteria.
The three-lepton plus $\MET$ category selects, in addition to the $Z$ boson, 
a third charged lepton with $E_T > 20$ GeV ($p_T > 20$ GeV/$c$ for muons)
and $\MET > 20$ GeV. For two-lepton events, in order to reduce the 
significant contribution from $WW$ boson pairs, we select $Z$ bosons using a narrower 
invariant mass range of $86 < M_{ll} < 96\ {\rm GeV}/c^2$. 
Two additional requirements are designed to suppress the Drell-Yan background in the two-lepton 
category. The first requires $\MET$ significance ($\MET^{\mathrm sig}$) to be larger than 3 
GeV$^{\mathrm 1/2}$, where $\MET^{\mathrm sig}$ is defined as $\MET/\sqrt{\sum E_T}$ 
and the sum is over all calorimeter towers above a given threshold. 
If muons are identified in the event, $\sum E_T$ is also corrected for the muon momenta.
We find that $\MET^{\mathrm sig}$ is a better discriminant than $\MET$ in controlling the 
Drell-Yan background, and that the maximum expected signal significance is achieved when 
$\MET^{\mathrm sig}$ is at least 3 GeV$^{\mathrm 1/2}$.
The second requirement is for $\Delta\phi$ between $\MET$ and the closest lepton or jet, 
$\Delta\phi$($\MET$, lepton/jet), to be larger than 20$^\circ$, in order to reduce the 
likelihood of falsely--reconstructed large $\MET$ due to mismeasured jets or leptons.
Finally, in the two-lepton category we only consider events with zero or one jet
to suppress $t\bar{t}$ background. In this analysis, jets are reconstructed using a cone of fixed radius 
$\Delta R = 0.4$~\cite{jetAlg}, and are counted if $E_T >$ 15 GeV and $|\eta| <$ 2.5.

The main background in the four-lepton and three-lepton categories is from ``fake--lepton'' events, 
in which jets have been misidentified as leptons in $Z$/$W$\,+\,jets events. 
The backgrounds in the two-lepton category include $WW$, $t\bar{t}$, Drell-Yan, and fake--lepton
events.

For each of the three categories of events, the total efficiency for
accepting a $ZZ$ or $ZW$ event can be expressed as 
$\epsilon_{\mathrm total} = \epsilon_{\mathrm ID} \times \epsilon_{\mathrm trigger} \times \epsilon_{\mathrm geom-kin}$,
where $\epsilon_{\mathrm ID}$ is the efficiency for identifying the number of leptons appropriate for a 
given category, $\epsilon_{\mathrm trigger}$ is the efficiency for the event to pass the trigger
requirements, and $\epsilon_{\mathrm geom-kin}$ is the efficiency 
for the leptons to fall within the geometric acceptance of the detector and for the events to 
pass all kinematic requirements for each signature.
The total efficiencies times branching ratios are listed in Table~\ref{tab:signal}.
The lepton identification efficiencies are measured using $Z\rightarrow ee/\mu\mu$ data.
The $Z\rightarrow ee$ events are selected with one identified electron and a second deposition
of energy in the electromagnetic calorimeter and an associated track.  
The $Z\rightarrow \mu\mu$ events are selected with one identified muon 
and a second track. The lepton pairs are required to have an invariant mass consistent with a
$Z$ and tracks of opposite charge. The unbiased lepton is used to measure the identification efficiency.
The trigger efficiencies are measured using data from independent trigger paths. The 
geometric and kinematic efficiencies are determined using a {\sc pythia} event 
generator~\cite{pythia} with a {\sc geant}-based detector simulation~\cite{geant}. 
Table~\ref{tab:signal} also shows the expected numbers of $ZZ$ and $ZW$ events, each
calculated using $\sigma \times \epsilon_{\mathrm total} \times \int {\mathcal{L}} d t$, where $\sigma$ is
the aforementioned NLO theoretical cross section.
In each of the three categories a relatively small, but non-negligible,
fraction of the total efficiency is due to final--state tau leptons
decaying to electrons and muons.
Overall, we expect 2.31 $\pm$ 0.29 $ZZ$ plus $ZW$ events in 
$(194 \pm 12)\ {\rm pb}^{\mathrm -1}$ of data.

\begin{table*}[htbp]
 \begin{center}
  \caption{The expected contributions from SM $ZZ$, $ZW$ and background sources 
	in 194 pb$^{\mathrm -1}$, and the observed number of candidates in the data. 
	The parentheses show the total efficiency times branching ratio for
	accepting $ZZ$ or $ZW$ events. Systematic and statistical uncertainties, 
	and the uncertainties of the $ZZ$ and $ZW$ NLO cross sections, are included.}
    \vspace*{0.3cm}
    \begin{tabular}{ccccc} 
    Process                 & 4 leptons       & 3 leptons        & 2 leptons       &  Combined \\ \hline
    $ZZ$	            & 0.06 $\pm$ 0.01 & 0.13 $\pm$ 0.01  & 0.69 $\pm$ 0.11 & 0.88  $\pm$ 0.13 \\
    ($\epsilon_{\mathrm zz}$ $\times$ 10$^3$)       & (0.22)           & (0.48)          & (2.56)            & (3.26) \\
    $ZW$                    & --              & 0.78 $\pm$ 0.06  & 0.65 $\pm$ 0.10 & 1.43  $\pm$ 0.16 \\
    ($\epsilon_{\mathrm zw}$ $\times$ 10$^3$)       & (--)             & (1.10)          & (0.92)            & (2.02) \\ \hline
    Total Signal            & 0.06 $\pm$ 0.01 & 0.91 $\pm$ 0.07  & 1.34 $\pm$ 0.21 & 2.31  $\pm$ 0.29 \\ \hline

    $WW$                    & --               & --                & 0.40 $\pm$ 0.07 & 0.40  $\pm$ 0.07 \\
    Fake                    & 0.01 $\pm$ 0.02 & 0.07 $\pm$ 0.06  & 0.21 $\pm$ 0.12 & 0.29  $\pm$ 0.16 \\
    Drell-Yan               & --               & --                & 0.31 $\pm$ 0.17 & 0.31  $\pm$ 0.17 \\
    \ttbar                  & --               & --                & 0.02 $\pm$ 0.01 & 0.02  $\pm$ 0.01 \\ \hline
    Total Background        & 0.01 $\pm$ 0.02 & 0.07 $\pm$ 0.06  & 0.94 $\pm$ 0.22 & 1.02  $\pm$ 0.24 \\ \hline

    Signal + Background        & 0.07 $\pm$ 0.02 & 0.98 $\pm$ 0.09  & 2.28 $\pm$ 0.35 & 3.33 $\pm$ 0.42 \\ \hline
    Data                              & 0 & 0 & 3 & 3                \\
    \end{tabular} 
    \label{tab:signal}
  \end{center}
\end{table*}  

The systematic uncertainties associated with the signal acceptances are dominated
by the Monte Carlo simulation of $\MET^{\mathrm sig}$ in the two-lepton category. 
This uncertainty is estimated by comparing distributions of $\MET^{\mathrm sig}$ between data 
and Monte Carlo in inclusive $W$ events,
where neutrinos from the $W$ decays produce large $\MET^{\mathrm sig}$. 
Relative to the signal acceptance, the uncertainty due to $\MET^{\mathrm sig}$ is 10\% for $ZZ$ and 
6\% for $ZW$. 
The other systematic uncertainties include those from lepton identification efficiencies 
(1\%), trigger efficiencies (1\%), the efficiency of the zero or one jet requirement (2\%),
dependence on different PDF's (2\%), and the calorimeter energy scale and resolution (3\%).
The total uncertainty in the efficiency estimate is 11\% for $ZZ$ and 8\% for $ZW$.

Backgrounds to the $ZZ$ and $ZW$ events are determined using a combination of data
and Monte Carlo simulations. The $WW$, Drell-Yan, and $t\bar t$ background estimates 
are obtained using {\sc pythia} Monte Carlo 
samples with the expected numbers of events normalized to the theoretical cross sections: 
12.4 pb for $WW$ from the MCFM program, 330 pb for Drell-Yan from {\sc pythia} 
($M(\gamma^*/Z^0) > 30$ GeV/c$^2$ and 
including a $K$--factor of 1.4~\cite{kfactor}), and 7 pb for $t\bar t$~\cite{ttbar}. 
A systematic uncertainty of 14\% on the 
$WW$ background results from the same effects that lead to
the uncertainties on the $ZZ$ and $ZW$ acceptances.
The Drell-Yan background has a 50\% uncertainty from two main sources: 35\% from the modeling 
of $\MET^{\mathrm sig}$, and 
another comparable amount from $\Delta\phi(\MET$, lepton/jet). 
The first is estimated from the comparison of 
$\MET^{\mathrm sig}$ distributions between Drell-Yan data and Monte Carlo, and the
second from the observed change in efficiency of the $\Delta\phi(\MET$, lepton/jet) requirement
after adjusting the jet energy scale.
The \ttbar\ background has a 15\% uncertainty due primarily to the uncertainty in the
jet energy scale. 

The fake--lepton background is obtained entirely using data. 
First, the probability for a jet to be misidentified as an electron or muon (fake rate) 
is estimated from jet-triggered
data samples after subtracting real leptons from $W$ and $Z$ decays. The lepton fake rates are averaged
over four samples with increasingly harder jet $E_T$ spectra. 
The observed differences between the jet samples are used to estimate the 
uncertainties in the lepton fake rates.
The fake--lepton background is then determined by applying these fake rates to jets in 
lepton-triggered events which would have passed the event selection had one jet faked a lepton. 
This background has a 41\% uncertainty, dominated by the uncertainty associated with the lepton
fake rates. The backgrounds in the three event categories are summarized in Table~\ref{tab:signal}.

\begin{figure*}[htbp]
  \begin{center}
   \mbox{\psfig{figure=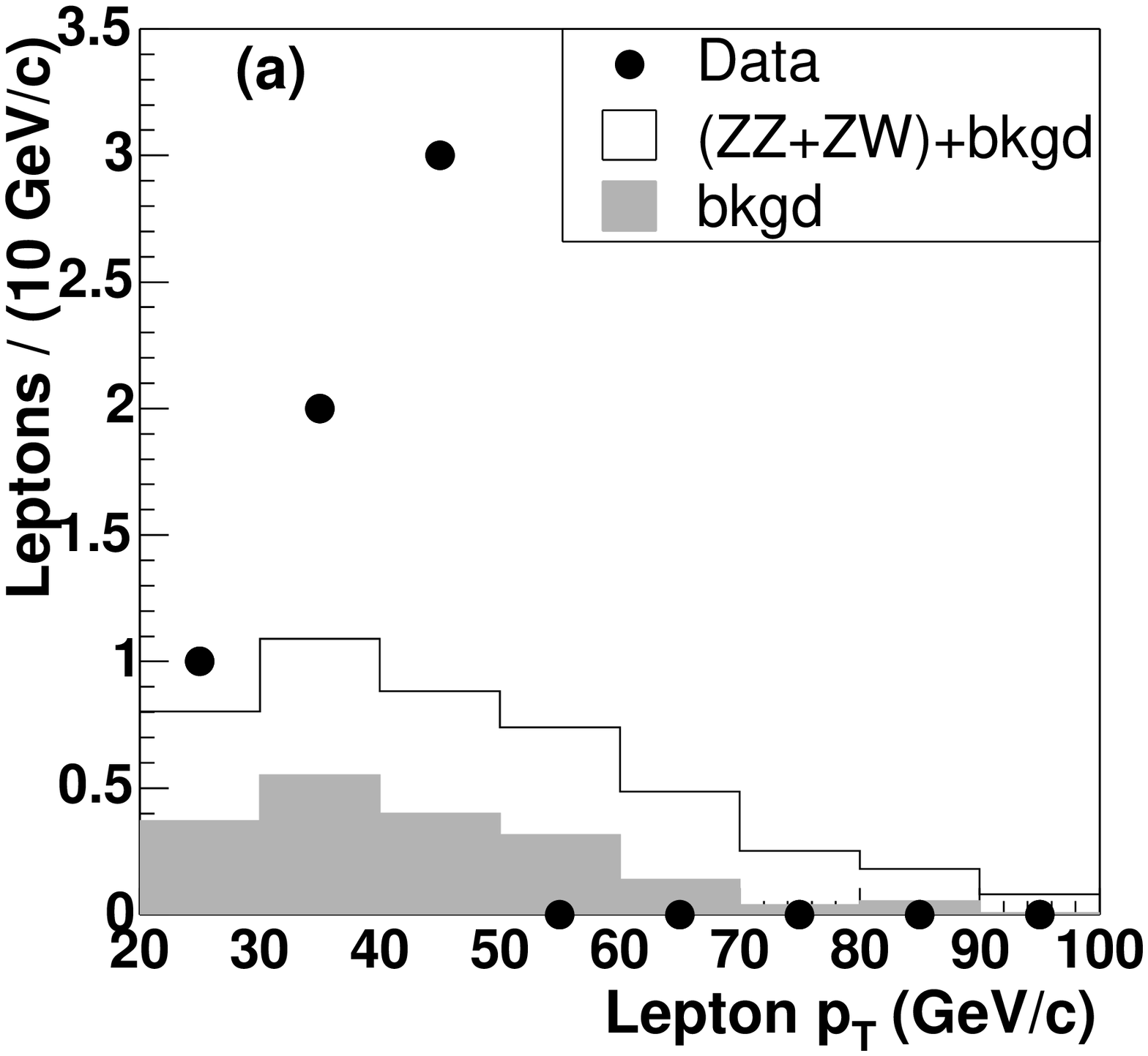,width=8.0cm} 
	\psfig{figure=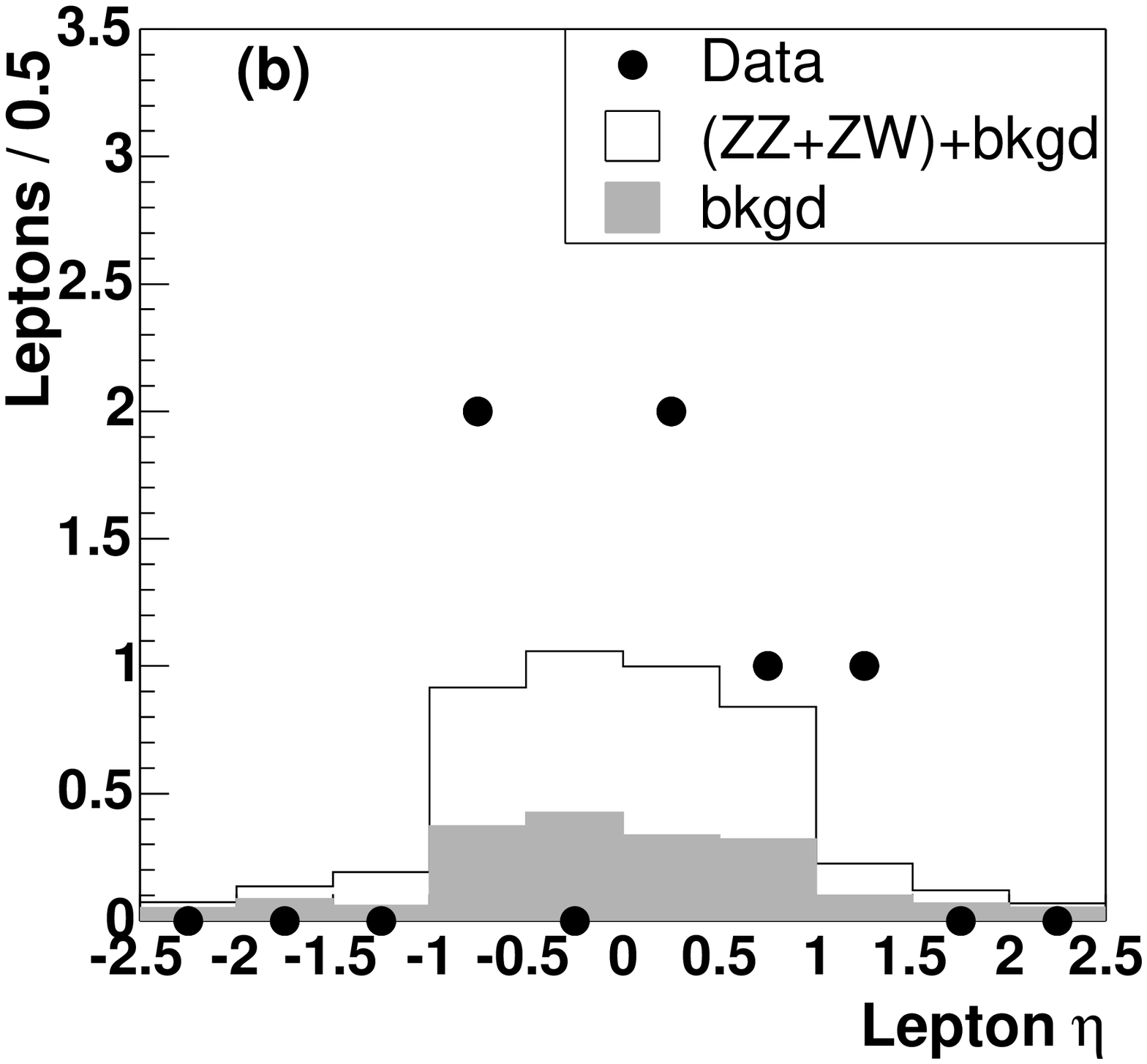,width=8.0cm}} \\
    \caption{Distributions of (a) lepton $p_T$ and (b) lepton $\eta$ of the candidate data events, and 
     the expected SM contributions in the two--lepton plus $\MET$ category.}
    \label{fig:lep}
  \end{center}
\end{figure*}

After all selection criteria we observe 3 events in the data\footnote{Another event passes all
the requirements for $ZZ\rightarrow eeee$ in the four--lepton category, except for
an isolation cut on one electron~\cite{estar}.}, all of them in the two-lepton plus $\MET$ category
(2 $ee$ and 1 $\mu\mu$), compared to a SM signal plus background 
expectation of $3.33 \pm 0.42$ events.
In Figure~\ref{fig:lep} we present distributions of lepton $p_T$ 
and $\eta$ in the two-lepton plus $\MET$ category, comparing data to SM expectations.
The probability for the background of $1.02\pm 0.24$ events to fluctuate to 
give three or more events is 9\%. Therefore, we set a 95\% confidence level 
upper limit on the $ZZ$ and $ZW$ combined cross section by applying a Bayesian 
method~\cite{bayes} with a flat prior cross section probability above zero.
Using the Poisson statistics for the data and including the assumed Gaussian 
uncertainties in the expected signal and background, we find that
the $ZZ$ and $ZW$ combined cross section is less than 15.2 pb at the 95\% confidence level. 
Using this analysis we conclude that
about 1 fb$^{\mathrm -1}$ of data is needed for a 3$\sigma$ measurement 
of the SM cross section for $ZZ+ZW$ production.

In summary, a search for $ZZ$ and $ZW$ production in $p\bar p$
collisions at $\sqrt s$ = 1.96 TeV has been performed using the leptonic
decays of the vector bosons. 
In a data sample corresponding to an integrated luminosity of 194 pb$^{\mathrm -1}$, 3
candidates are found with an expected background of 1.02  $\pm$ 0.24 events.
The predicted number of $ZZ$ and $ZW$ events is 2.31 $\pm$ 0.29. 
A 95\% confidence level limit on the sum of the production 
cross sections for $p\bar p \rightarrow ZZ$ and $p\bar p \rightarrow ZW$ is measured to be 
15.2 pb, consistent with the standard model prediction of $5.0 \pm 0.4$ pb.

We thank the Fermilab staff and the technical staffs of the participating institutions for their vital contributions. This work was supported by the U.S. Department of Energy and National Science Foundation; the Italian Istituto Nazionale di Fisica Nucleare; the Ministry of Education, Culture, Sports, Science and Technology of Japan; the Natural Sciences and Engineering Research Council of Canada; the National Science Council of the Republic of China; the Swiss National Science Foundation; the A.P. Sloan Foundation; the Bundesministerium fu\"r Bildung und Forschung, Germany; the Korean Science and Engineering Foundation and the Korean Research Foundation; the Particle Physics and Astronomy Research Council and the Royal Society, UK; the Russian Foundation for Basic Research; the Comision Interministerial de Ciencia y Tecnologia, Spain; and in part by the European Community's Human Potential Programme under contract HPRN-CT-2002-00292, Probe for New Physics.


\begin{thebibliography}{99}
\bibitem{xyz}
CDF uses a ($z$, $\phi$, $\theta$) coordinate system, where the $z$-axis is in the direction
of the proton beam, and $\phi$ and $\theta$ are the azimuthal and polar angles respectively. 
The pseudorapidity $\eta$ is defined as $-\ln (\tan {\frac {\theta} {2}})$.
The transverse momentum of a charged particle is $p_T = p\sin \theta$, where $p$ represents 
the measured momentum of the track. The analogous quantity using
calorimeter energies, defined as $E_T = E\sin \theta$, is called transverse energy.
\end{thebibliography}
\end{document}